\documentclass[aip,jcp,
  asmath,amssymb,
  preprint,groupedaddress]{revtex4}

\usepackage{graphicx}
\usepackage{epstopdf}
\usepackage{amssymb, amsmath}
\usepackage{multirow}
\usepackage{dcolumn}
\usepackage{bm}
\usepackage{subfigure}
\usepackage{color}
\usepackage{ulem}

\usepackage[final]{hyperref}
\hypersetup{
        colorlinks=true,       
        linkcolor=blue,          
        citecolor=blue,        
        filecolor=magenta,      
        urlcolor=blue
        }

\usepackage[utf8]{inputenc}
\usepackage[T1]{fontenc}
\usepackage{mathptmx}
\usepackage{etoolbox}

\usepackage{xr}
\newcommand\ket[1]     {|{{#1}}\rangle}
\newcommand\bra[1]     {\langle{{#1}}|}

\newcommand\PsiGS      {\Psi_0}

\newcommand\Hop        {{\hat{H}}}
\newcommand\Kop        {{\hat{K}}}

\newcommand\Vop        {{\hat{V}}}

\newcommand\Order[1]   {\mathcal{O}\left(#1\right)}

\newcommand\eql[2] 
{
\begin{equation}\label{#1}
\begin{split}
#2
\end{split}
\end{equation}
}

\newcommand\eqsl[1]                            
{
\begin{align}
#1
\end{align}
}

\makeatletter
\def\@email#1#2{%
 \endgroup
 \patchcmd{\titleblock@produce}
  {\frontmatter@RRAPformat}
  {\frontmatter@RRAPformat{\produce@RRAP{*#1\href{mailto:#2}{#2}}}\frontmatter@RRAPformat}
  {}{}
}%
\makeatother

\begin{document}
 
\title{Ab initio Calculations in Atoms, Molecules, and Solids,
Treating Spin-Orbit Coupling and Electron Interaction on Equal Footing}
 
\author{Brandon Eskridge}
\email{bkeskridge@wm.edu}
\affiliation{Department of Physics, College of William and Mary, Williamsburg, Virginia 23187, USA}

\author{Henry Krakauer}
\affiliation{Department of Physics, College of William and Mary, Williamsburg, Virginia 23187, USA}

\author{Hao Shi}
\affiliation{Department of Physics and Astronomy, University of Delaware, Newark, Delaware 19716, USA} 

\author{Shiwei Zhang}
\affiliation{Center for Computational Quantum Physics, Flatiron Institute, New York, NY 10010, USA}

\begin{abstract}

We incorporate explicit, non-perturbative treatment of spin-orbit coupling into {\it ab initio} auxiliary-field quantum Monte Carlo (AFQMC) calculations. 
The approach allows a general computational framework for molecular and bulk systems in which materials specificity, electron correlation, and spin-orbit 
coupling effects can be captured accurately and on equal footing, with favorable computational scaling versus system size. We adopt relativistic effective-core potentials which 
have been obtained by fitting to fully relativistic data and which have demonstrated a high degree of reliability and transferability in molecular systems.
This results in a 2-component spin-coupled Hamiltonian, which is then treated by generalizing the {\it ab initio} AFQMC approach. We 
demonstrate  the method by computing the electron affinity in Pb,  the bond dissociation energy in Br$_2$ and I$_2$, and solid Bi. 

\end{abstract}

\maketitle

\section{Introduction}

Understanding the interplay between electron-electron interactions and spin-orbit coupling (SOC) in molecules and materials has been a long standing goal in the fields of physics and chemistry.
This goal has become especially important with the relatively recent advances in broad classes of topological materials, and the significant technological potentials of the interplay between SOC and interaction effects. 
The solution of the many-body problem including SOC is central to achieving this understanding.
Typically, the electron interaction effects are much larger than the SOC effects.
Perturbative inclusion of SOC is, of course, possible and produces accurate results for light atoms, where SOC is weak, and for systems in which electron correlation and SOC are only weakly coupled~\cite{MALMQVIST2002230}.
However,
 perturbative treatments, which 
 neglect any ``feedback'' effects of SOC on electron correlation,
 are often inadequate for 
 capturing the interplay between the two, and incapable of providing predictive calculations in the many systems of interest.
In such cases, 
SOC and electron-electron interactions both must be treated on equal footing at the many-body level of theory.

Solving the many-body problem is highly challenging, with computational costs which in general grow exponentially with system size.
Explicit inclusion of SOC expands the dimension of the Hilbert space and makes the problem even more challenging to treat. 
Approximate solutions based on density functional theory (DFT) which include 
SOC are possible and are able to produce accurate results in a wide variety of problems;
 however, just as for the case of non-spin-dependent Hamiltonians (which we will refer to as 
  spin-free for brevity), DFT may be inadequate in strongly correlated systems.
 Many-body methods from quantum chemistry have been generalized to SOC, and provide a very good approach for molecular systems~\cite{Chan2016,Sharma2018,Hess1998,SOCCSD_2018}.
Because of the
steeper computational scaling, however, extensions to larger  systems and solids are not straightforward.
There has also been work to extend diffusion Monte Carlo to treat SOC, in which the spin flip terms are sampled stochastically~\cite{Mitas2016}.
 
In this work, we generalize the auxiliary-field quantum Monte Carlo (AFQMC) method for the treatment of SOC in \textit{ab initio} calculations in molecules and solids.
AFQMC works directly in Slater determinant space; it shares much of the same machinery as DFT and can treat the same Hamiltonian as standard post-DFT methods from quantum chemistry. 
SOC effects in the two-component form can be treated exactly in AFQMC.
We develop and implement this feature and combine it with the standard \textit{ab initio} spin-free phaseless AFQMC  technologies \cite{Zhang2003,AlSaidi2006b,Suewattana2007,Purwanto2008,Purwanto2009_C2,Purwanto2013,Motta2018} to obtain a general method which can treat SOC and electron correlation effects on equal footing. 
The high accuracy of AFQMC, combined with its gentle scaling with system size compared to other many-body methods, makes this approach an excellent 
tool for the study of the interplay between SOC and electron-electron interaction.

The most direct treatment of spin-orbit coupling is to use a relativistic Hamiltonian which, of course, includes scalar relativistic effects as well.
The most complete relativistic Hamiltonian for quantum chemistry problems is given by the Dirac-Coulomb-Breit (DCB) Hamiltonian in the no-pair approximation which has 4-component (4c) bispinor solutions, as we discuss 
further in Section~\ref{sec:2cHam}~\cite{Dolg2012}.
The  4c-bispinors consist of the so-called large and small component spinors, which each have two components, and together they describe both electronic and positronic degrees of freedom.
For the study of molecules and materials, we are interested in the electronic solutions only, but the local potential of the atomic nuclei in the system leads to a mixing of the small and large components.
While it is possible to directly treat the 4c-DCB Hamiltonian in the no-pair approximation, as has been done with several methods, 
the inclusion of 4 components greatly increases the system size at the many-body level of theory.
To address this bottleneck, the large and small components can be decoupled via an appropriate transformation to arrive at a two-component (2c) Hamiltonian.
While there are many methods for doing this, some of the most popular in the literature are the Breit-Pauli~\cite{Pauli1927,Breit1932}, Douglas-Kroll-Hess~\cite{Hess1985,Hess1986}, and exact two-component (x2c)~\cite{X2cDyall, X2c2005, X2C2007a, X2C2007b} methods among others.

The most dramatic relativistic effects occur for deep core electrons which move at an appreciable fraction of the speed of light in heavy atoms.
Much of the relativistic effects experienced by the valence electrons are due to the indirect effects caused by the modification of the core.
Because of this, 
it is common practice to use two-component
relativistic effective core potentials (REPs) / pseudopotentials (PPs) which have been fit to fully relativistic data. 
REPs/PPs have the advantage of bypassing much of the complexity and subtlety involved in using the 
all-electron,
two-component Hamiltonians as well as reducing the system size in many-body calculations.
Furthermore, they have been  demonstrated to produce accurate
results
when combined with an appropriate many-body method,
as shown via direct comparisons with experiment.
REPs/PPs are often published  along with systematically convergent basis sets which allow an extrapolation to the complete basis set (CBS) limit to be performed.

The AFQMC method  has demonstrated a high degree of accuracy in both molecules and materials, as well as in lattice models of strongly correlated materials such as the Hubbard model.
AFQMC has recently been extended to include SOC and demonstrated in
model systems~\cite{Rosenberg2019,Rosenberg2017,Rosenberg2016}.
In the present work, we extend the adaptation of AFQMC including SOC directly to realistic \textit{ab initio} Hamiltonians.
We focus on two-component REPs/PPs which explicitly include the spin-orbit coupling operator as opposed to two-component relativistic all-electron Hamiltonians;
 however, all-electron Hamiltonians can be used in the formalism presented here with no modifications at the AFQMC level of theory.

The remainder of the paper is organized as follows.
In Section~\ref{sec:Method}, we provide theoretical background on the relativistic quantum many-body problem,
 as well as a brief overview of AFQMC for \textit{ab initio} calculations.
We also detail the modifications to the typical AFQMC formulation to include SOC.
In Section~\ref{sec:Apps}, we describe several test applications of the ab initio AFQMC approach including SOC, 
compared with corresponding scalar relativistic calculations.
We begin with the calculation of the electron affinity (EA) of lead, followed by the dissociation energy 
of heavy-elements dimers.
Next, we compute the cohesive energy and equation of state of crystalline Bi.
We conclude with some general remarks in Section~\ref{sec:Discussion}.

\section{Theory}
\label{sec:Method}

In this section, we describe the treatment
of relativistic effects with 
AFQMC, including spin-orbit (SO) interactions.
Many-body AFQMC calculations for realistic molecular and condensed-matter Hamiltonians are formulated 
using second-quantization, expressed with respect to a chosen finite-basis set of one-particle orbitals, such as planewaves or
gaussian-type-orbitals (GTOs). Although AFQMC has favorable system size scaling ($N^3 - N^4$,  
where $N$ is the number of electrons), the computational demands grow rapidly for heavier atoms.
To avoid this, we adopt two-component fully relativistic PPs,
which have proved very accurate, even for heavy atoms, through comparison with 4-component all-electron calculations and with experiments.
Thus, spin and orbital degrees of freedom are included from the outset, 
avoiding post-processing  perturbative or second-variational SO treatments. 
Our relativistic AFQMC approach is therefore general and applicable to systems with arbitrary compositions 
of light and heavy constituent atoms. 

%
%

\subsection{Relativistic many-body problem}
\label{sec:2cHam}

While quantum field theory provides a fully relativistic treatment of quantum electrodynamics (QED), it is not feasable for all but the smallest problems. Practical relativistic many-body treatments usually start from the 4-component Dirac-Coulomb-Breit (DCB) electronic Hamiltonian \cite{EsRelQChem,Rev_Autschbach2012,Dolg2012}
\eql{eq:dcb}
{
\hat{H}_\mathrm{DCB}
 =  \hat{H}_\mathrm{Dirac}+ \hat{H}_\mathrm{Int}, 
}
where 
\eql{eq:dcb1}
{
\hat{H}_\mathrm{Dirac} = \sum_i \left ( {\bm{\alpha_i}} \cdot \vec{p}_i +  \bm{\beta_i} c^2 + \sum_A \hat{V}^A (r_{iA}) \right )
}
is the one-body Dirac Hamiltonian, ${\bm{\alpha}}$ and $ \bm{\beta}$ are the usual Dirac matrices,
is the electron-nucleus interaction, and $i$ runs over all electrons.
The interaction Hamiltonian is given by 
\eql{eq:dcb2}
{
\hat{H}_\mathrm{Int} =\frac{1}{2} \sum_{i \neq j} \left (
\frac{1}{r_{ij}} -  \frac{1}{2 r_{ij}} \left[ \bm{\alpha}_i \cdot \bm{\alpha}_j + \frac{(\bm{\alpha}_i \cdot  r_{ij}) (\bm{\alpha}_j 
\cdot  r_{ij})  }{ r_{ij}^2} \right] \right ) 
\,,
}
where the first term is the usual electron-electron Coulomb repulsion, and the second term is the Breit interaction which makes small, but non-negligible, contributions.

The technical problems with the DCB Hamiltonian are well known. The Dirac Hamiltonian 
is not bound from below, so care must be used in variational calculations to prevent collapse to negative energy states. 
These can be addressed through the use of positive-energy projection operators in solving the DCB equations.
Residual issues still remain, however. For example, should the projection operators be optimized at the mean-field or 
correlated level of theory \cite{Almoukhalalati2016}. The 4-component DCB equations can also be reduced to two-component Hamiltonian formulations, 
where the negative-energy degrees of freedom are explicitly transformed away, 
as in the Breit-Pauli Hamiltonian and the zeroth-order regular approximation (ZORA) \cite{EsRelQChem}. 
Due to their relative simplicity, the two-component formulations are most commonly used in practical all-electron applications.
 
In heavy atoms, the core electrons experience strong direct relativistic effects while the valence electrons experience only weak direct relativistic effects with indirect effects due to the altered core dominating~\cite{Dolg2012}.
In the nonrelativistic limit, the Dirac Hamiltonian given in Eq.~\ref{eq:dcb1} can be trivially reduced to a one-component form if spin-dependent terms are neglected, or a two-component form if the lowest order spin-dependent terms are retained.
The former is the limit within which standard nonrelativistic electronic structure methods operate.
Similar to nonrelativistic treatments, the valence space is the most relevant for chemical and condensed matter systems.
This motivates the use of a valence-only model Hamiltonian, based on two-component PPs,
which operates
on a space with core degrees of freedom eliminated, given by
\eql{eq:relPP}
{
\hat{H} = \sum_i^{N_v} \left (  - \frac{1}{2}\nabla_i^2 + \sum_A \hat{V}_A^\mathrm{PP}(i) \right )+ \frac{1}{2}  \sum_{i \neq j}^{N_v} \frac{1}{r_{ij}} + \hat{V}_\mathrm{ion-ion} ,
}
where $N_v$ is the number of valence electrons, $\hat{V}_A^\mathrm{PP}$ is the relativistic PP corresponding to ion center $A$, and $\hat{V}_\mathrm{ion-ion}$ is the point-charge Coulomb ion-ion interaction (for non-overlapping cores).
The pseudopotential $\hat{V}_A^\mathrm{PP}$ is parameterized and fit such that it includes $\hat{V}_{A}$ from Eq.~\ref{eq:dcb1},
the core-valence electron interactions via the Coulomb interaction of Eq.~\ref{eq:dcb2},
and scalar relativistic corrections from the first two terms of Eq.~\ref{eq:dcb1}.
If the reference data includes the Breit interaction, both the core-valence and valence-valence Breit interaction is approximately included in $\hat{V}_A^\mathrm{PP}$.
Other relativistic effects, such as finite-nucleus effects and QED corrections, can be easily incorporated if suitable reference data is used to fit the 
PP~\cite{Dolg2012}.
It is common practice to separate the fully relativistic PP of Eq.~\ref{eq:relPP} into a scalar relativistic PP and a SO-PP as 
\eql{eq:relPP_sep}
{
\hat{V}_A^\textrm{PP}(i) = \hat{V}_A^{scalar-PP}(i) + \hat{V}^{SO-PP}_A(i)
\,,
}
where the scalar relativistic part, $\hat{V}_A^{scalar-PP}$, is constructed by averaging over the total angular momentum and the SOC part, $\hat{V}^{SO-PP}_A$, is implicitly defined.
This formalism facilitates the selective inclusion of either scalar relativistic effects only or both scalar and SO effects for a fully relativistic treatment, by simply including or excluding $\hat{V}_A^\textrm{SO-PP}$, as desired, for each ion center $A$.

Relativistic PPs offer a high degree of accuracy (see e.g.,~Ref.~\cite{Dolg2012}) and computational convenience. 
Compared to  direct solution of the 2- or 4-component all-electron equations, 
the computational cost is reduced by limiting the number of electrons which are explicitly treated. 
Systematically convergent basis sets are typically available for relativistic PPs, which is highly desirable for explicitly correlated many-body calculations, since it 
facilitates extrapolation to the CBS limit, 
whereas some 2-component methods require a reasonably complete basis to be used from
the outset.
For example, the so-called exact two-component methods~\cite{X2cDyall, X2c2005, X2C2007a,X2C2007b} perform a formally exact decoupling of the large and small components of the DCB Hamiltonian
via an algebraic transformation which requires a faithful matrix representation of the 4c-DCB Hamiltonian.
While this method has proven to be quite accurate compared with experiment, it complicates the typical CBS extrapolation scheme.
Use of relativistic PPs allows straightforward inclusion of relativistic effects in systems which contain both light and heavy atoms which is more
challenging in some 2-component all-electron methods.
Finally, relativistic PPs are straightforward to incorporate in existing methods since they are based on a non-relativistic form of the Hamiltonian.

%
%

\subsection{Auxiliary-Field Quantum Monte Carlo (AFQMC)}
\label{sec:AFQMC}

Here we provide a brief overview of the Auxiliary-Field Quanutm Monte Carlo (AFQMC) method~\cite{Zhang2003}.
The formalism as applied to molecular systems is presented in a recent review~\cite{Motta2018}.
In the next section  
we describe in further detail our adaptation of AFQMC to include SOC.

AFQMC is an orbitally-based, explicitly correlated electronic structure
method which uses the second quantized form of the many-electron Hamiltonian,
\eql{eq:2ndQuantHspat}
{
\Hop = \Kop + \Vop =  \sum_{\mu \nu} K_{\mu \nu} \hat{c}^{\dagger}_\mu  \hat{c}_\nu +  \sum_{ \mu \nu \gamma \delta } V_{\mu \nu \gamma \delta} \hat{c}^{\dagger}_\mu  \hat{c}^{\dagger}_\nu \hat{c}_\delta \hat{c}_\gamma\,,
}
where $\hat{c}^{\dagger}_\mu$ and $\hat{c}_\mu$ are fermionic creation and annihilation operators, respectively, corresponding to a
chosen orthonormal single-particle orbital basis which may consist of spatial orbitals, spin orbitals, planewaves, etc.
$\Kop$ includes all one-body Hamiltonian terms such as the electron kinetic energy, and the relativistic PPs (Eq.~\ref{eq:relPP}) described in Sec.~\ref{sec:2cHam},
and $\Vop$ includes all two-electron Hamiltonian terms.
All standard forms of the many-electron Hamiltonian can be expressed as in Eq.~\ref{eq:2ndQuantHspat} including nonrelativistic, relativistic, all-electron and psuedopotential treatments.

AFQMC uses imaginary time projection to obtain the many-body ground state beginning with some initial wavefunction which has non-zero overlap with the exact many-body ground state.
Formally, the projection of the many-body ground state, $ \ket{\PsiGS}$, is performed as follows,
\eql{eq:gs-proj}
{
    \lim_{\beta \to \infty}  e^{-\beta \Hop} \ket{\Psi_I} = 
    e^{-\tau \Hop}
    e^{-\tau \Hop}
    \cdots
    e^{-\tau \Hop}
    \ket{\Psi_I}
    \to
    \ket{\PsiGS}
    \,
}
where the total projection time, $\beta$, has been divided into small imaginary time steps, $\tau$,  and $\ket{\Psi_I}$ is an 
initial wavefunction.

The imaginary time projection operator is applied by utilizing the fact that
the exponential of a one-body operator acting upon a Slater determinant simply produces another Slater determinant~\cite{THOULESS1960225}; 
however, the presence of two-body Hamiltonian terms makes the projection nontrivial.
To perform the projection in practice, $e^{-\tau \hat{H}}$ is mapped onto a product of exponentials of one-body operators as follows.
First, the Trotter-Suzuki decomposition~\cite{Trotter1959,Suzuki1976} is used,
\eql{eq:Trotter-Suzuki}
{
    e^{-\tau \Hop}
    \approx
    e^{-\tau \Kop / 2}
    e^{-\tau \Vop}
    e^{-\tau \Kop / 2}
    + \Order{\tau^3}
    \,,
}
which separates the exponential operator of explicit one-body terms, $\Kop$, 
from the two-body terms, $\Vop$, at the cost of introducing a finite Trotter step error which can be systematically removed by extrapolation to $\tau = 0$.
As for the operator $e^{-\tau \Vop}$, the two-body Hamiltonian term can be rewritten as a quadratic form of one-body operators,
\eql{eq:Vquad}
{
\Vop = \sum_\gamma \hat{v}^2_\gamma 
}
where, for example, $\hat{v}_\gamma$ could be obtained from a Cholesky decomposition~\cite{Purwanto2011}, from density fitting, or some other method.
Typically, the set of $\hat{v}_\gamma$ is truncated based on a convergence threshold, where
the errors can be systematically removed, if needed.
The factored from of $\Vop$ allows the Hubbard–Stratonovich~\cite{Stratonovich1957,Hubbard1959} transformation to be applied,
\eql{eq:HS1}
{
    e^{-\tau \Vop}
    =
    \int d\bm{\sigma} P(\bm{\sigma})
\:e^{\sqrt{\tau} \bm{\sigma} \cdot {\hat{\mathbf v}}}
    \,,
}
where $\bm{\sigma}$ is a vector containing auxiliary-fields, $\hat{\mathbf v}$ is the vector of one-body operators defined in Eq.~\ref{eq:Vquad}, and $P$ is a normal distribution function.
Combing Eqs.~\ref{eq:Trotter-Suzuki} and~\ref{eq:HS1} ,
\eql{eq:HSfull}
{
    e^{-\tau \Hop}
    \approx
      \int d\bm{\sigma} P(\bm{\sigma}) B(\bm{\sigma})
\,,
}
where $B(\bm{\sigma}) =   e^{-\tau \Kop / 2} e^{\sqrt{\tau} \bm{\sigma} \cdot {\hat{\mathbf v}}}  e^{-\tau \Kop / 2} $ involves only exponentials of one-body operators.

The projection of Eq.~\ref{eq:gs-proj} has now been mapped onto a high-dimensional integral over many auxiliary-fields with an integrand which can be easily applied to individual Slater determinants.
Eq.~\ref{eq:HSfull} is evaluated via stochastic sampling as a branching random walk in non-orthogonal Slater determinant space with auxiliary fields, $\bm{\sigma}$, sampled from $P(\bm{\sigma})$.
At each imaginary time step a single Slater determinant in the many-body wavefunction is individually updated as,
\eql{eq:walkUpdate}
{
\ket{\Phi^{(i)}_k} = B(\bm{\sigma}) \ket{\Phi^{(i-1)}_{k}}
\,.
}
where $i$ is the index of the projection step, $\ket{\Phi^{(i)}_k}$ is a single Slater determinant random walker, and the many-body ground state has a Monte Carlo representation,
\eql{eq:mcPsi0}
{
\ket{\Psi_0} \doteq \sum_k \ket{\Phi_k}
\,.
}

The above reformulation of the ground-state projection as a branching random walk \cite{CPMC-PRL95,CPMC-PRB97,Zhang2003}, which is 
sometimes referred to as free-projection,  
allows the introduction of importance sampling and the control of the sign and phase problems with a sign or gauge condition in Slater determinant space.
For general electron-electron interactions, such as the Coulomb interaction,
an exponential growth in stochastic noise arises as the walk progresses.
This is a manifestation of the generic fermionic sign problem which arises in almost all fermionic quantum Monte Carlo methods.
The origin of the phase problem is in the fact that physical observables computed with the exact many-body wavefunction are invariant to an overall complex phase.
Thus, any wavefunction of the form $e^{-i\phi} \ket{\Psi_0}$ with $\phi \in (0, 2\pi]$  is an equally valid ground state.
As the walk progresses, the ensemble of random walkers can be sampled from any such wavefunction.
Since the set of operators $\hat{v}_\gamma$ are complex-valued in general, the orbitals within each random walker accumulate a random phase over the course of the projection,
causing the random walk to ``hop'' between different degenerate states with different gauge choices~\cite{Zhang2003}.

To control this problem, the phaseless approach~\cite{Zhang2003} first applies an importance 
sampling with importance function $\bra{\Psi_T} \Phi_k \rangle$.
Eq.~\ref{eq:walkUpdate} becomes
\eql{eq:walkUpdateImp}
{
\ket{\Phi^{(i)}_k} = W(\sigma, \Phi^{(i-1)}_{k}) B(\bm{\sigma} - \bar{\bm{\sigma}}) \ket{\Phi^{(i-1)}_{k}}
\,,
}
where W is given by,
\eql{eq:weightImp}
{
W(\sigma, \Phi^{(i)}_{k}) \equiv \frac{\bra{\Psi_T} B(\sigma - \bar{\bm{\sigma}}) \ket{\Phi^{(i)}_{k}}}{\bra{\Psi_T} \Phi^{(i)}_k \rangle}
 \mathrm{Exp}[\bm{\sigma}\cdot \bar{\bm{\sigma}} - (  \bar{\bm{\sigma}} \cdot \bar{\bm{\sigma}}/2) ]
\,,
}
and $\bar{\bm{\sigma}}$, known as the force bias, is given by,
\eql{eq:forceBias}
{
\bar{\bm{\sigma}} = - \sqrt{\tau} \frac{\bra{\Psi_T} \hat{\bm{v}} \ket{\Phi_k} }{\bra{\Psi_T} \Phi_k \rangle}
\,.
}
The Monte Carlo representation of the many-body wavefunction becomes,
\eql{eq:mcPsi0}
{
\ket{\Psi_0} \doteq \sum_k w_k \ket{\Phi_k}
\,,
}
where $w_k$ is the weight of $\ket{\Phi_{k}}$ which is accumulated over the course of the projection 
as $w^{(i)}_{k} = W(\bm{\sigma}, \Phi^{(i-1)}_{k})  w^{(i-1)}_{k}$.

Physical observables are computed during the projection using the Monte Carlo representation of the many-body wavefunction in Eq.~\ref{eq:mcPsi0}.
For example, the ground state energy is computed using the mixed-estimator,
\eql{eq:mixE1}
{
E_0 = \frac{\bra{\Psi_T} \hat{H} \ket{\Psi_0}}{\bra{\Psi_T} \Psi_0 \rangle} = \lim_{\beta \to \infty} \frac{\bra{\Psi_T} \hat{H}e^{-\beta \Hop} \ket{\Psi_I} }{\bra{\Psi_T} e^{-\beta \Hop} \ket{\Psi_I}}  
\,,
}
where $ \ket{\Psi_T} $ is a trial wave function which corresponds to an estimate of the exact ground state.
\eql{eq:mixE2}
{
E_0 \approx \frac{\sum_{k} w_k E_L \left[ \Phi^k \right] }{\sum_k w_k}
\,,
}
where $E_L$ is the local energy given by,
\eql{eq:localE}
{
E_L \left[ \Phi^k \right] = \frac{\bra{\Psi_T} \Hop \ket{\Phi^k}}{\bra{\Psi_T} \Phi^k \rangle}
\,.
}

A projection step follows importance sampling to eliminate the sign or phase problem. 
Since the random walkers are able to diffuse across the entire
complex plane defined by $\bra{\Psi_T} \Phi_k \rangle$,
a nonzero density of walkers accumulate at the complex origin, which
causes the weight of the walkers to fluctuate and diverge.
This is removed by a projection. 
At each step, walkers are individually projected onto an evolving line in the complex plane by multiplying the walker weights by max\{$0,\textrm{cos}(\Delta \theta)$\} where $\Delta \theta$ is the phase of $\bra{\Phi_T} \Phi_k \rangle$ /  $\bra{\Phi_T} \Phi_{k-1} \rangle$.

%
%

\subsection{Spin-orbit coupling in AFQMC}
\label{sec:socAFQMC}

In this section, we discuss in detail the treatment of SOC
in AFQMC for realistic Hamiltonians.
In principle, the formalism in Sec.~\ref{sec:AFQMC} is general and can include SOC if we treat $\mu\nu$ as spin-orbit indices \cite{Rosenberg2019}.
We outline this formalism in detail below in the context of \textit{ab initio} SOC Hamiltonians discussed in Sec.~\ref{sec:2cHam}
In a relativistic treatment, the one-electron part of the many-body Hamiltonian (Eq.~\ref{eq:2ndQuantHspat}) can be explicitly expanded as,
\eql{eq:explicitH}
{
\Kop = \hat{T} + \hat{W}^{scalar} +  \hat{\bm{W}}^{soc} \cdot \hat{\bm{S}}
\,,
}
where $\hat{T}$ is the kinetic energy,
$\hat{W}^{scalar}$ includes all other scalar relativistic terms, and
$\hat{\bm{W}}^{soc} \cdot \hat{\bm{S}}$ is the SOC operator where 
we have factored $\hat{V}^{SO-PP}_A = \hat{\bm{W}}^{soc} \cdot \bm{S}$ such that $\hat{\bm{W}}^{soc}$ is a vector with three spatial components.
The matrix elements with respect to spatial orbitals are given explicitly by
\eql{eq:WscalarElems}
{
\left[ W^{scalar} \right]_{ij} =  \int d\vec{r} \phi_i^*(\vec{r}) 
\sum_A {V_A^{scalar-PP}} 
\phi_j(\vec{r})
}
\eql{eq:WSocOpMatElems}
{
\left[W^{SOC}_\lambda\right]_{ij} = \int d\vec{r} \phi_i^*(\vec{r}) 
\sum_A W_{A,\lambda}^{soc}
\phi_j(\vec{r})\,,
}
where $\lambda = x,y,z$, and $i$ and $j$ are the spatial orbital indices associated with $\mu$, and $\nu$, respectively.
Other 2-component forms of the relativistic Hamiltonian, including all-electron treatments, can be expressed similarly to Eq.~\ref{eq:explicitH}.
In the discussion that follows, it is convenient to separate the SO-interaction into the axial part, which is parallel to the spin quantization axis, and the part which is transverse to the spin quantization axis, as
\eql{eq:WsocParPerp}
{
\hat{\bm{W}}^{soc} \cdot \hat{\bm{S}} = \hat{W}_{\parallel}^{soc} \hat{S}_{\parallel} + \hat{\bm{W}}_\bot^{soc} \cdot \hat{\bm{S}}_\bot,
}
where $ \hat{W}_{\parallel}^{soc} =  \hat{W}_z^{soc}$ and $\hat{\bm{W}}_\bot^{soc} =  \hat{W}_x^{soc} \hat{x}  +  i \hat{W}_y^{soc} \hat{y}$ if the $z$-axis is chosen as the quantization axis.

When the Hamiltonian has no explicit spin-dependence, i.e. in a scalar relativistic or nonrelativistic treatment, the Slater determinant random walkers are  factored into separate spin-sectors
\eql{eq:sdFac}
{
\ket{\Phi_k}= \ket{\Phi^\uparrow_k} \otimes \ket{\Phi^\downarrow_k}
\,,
}
where the $\uparrow$ and $\downarrow$ labels correspond to up and down spin, respectively, along some chosen quantization axis.
Similarly, the projection operator can be factored as,
\eql{eq:facBop}
{
B(\bm{\sigma}) = B^{\uparrow}(\bm{\sigma}) \otimes B^{ \downarrow}(\bm{\sigma})
\,.
}
If the charge decomposition is used, as has often been the case in electronic structure calculations, then
$B(\bm{\sigma})^{\uparrow} = B(\bm{\sigma})^{\downarrow}$.
Depending on the initial and trial wave function forms, 
each spin sector is either treated separately  using unrestricted determinants,
 or treated identically (hence saving half the propagation cost) as only one spin sector is explicitly operated upon.

When spin-orbit coupling is included via PP or otherwise,
the Hamiltonian in second-quantized form contains spin-flip terms.
The random walkers in AFQMC are
generalized Slater determinants.
In a basis of spin orbitals, $\chi^\sigma_i = \psi^\sigma_i (\vec{r}) \ket{\sigma}$, the walkers   
 $\ket{\Phi^G_k}$ have a block matrix representation given by
\eql{eq:gDet}
{
\Phi^G_k = 
  \begin{bmatrix}
   \Phi_k^{\uparrow \uparrow} &
   \Phi_k^{\uparrow \downarrow}  \\
    \Phi_k^{\downarrow \uparrow} &
   \Phi_k^{\downarrow \downarrow}
   \end{bmatrix}
\,.
}
The block matrices in Eq.~(\ref{eq:gDet}) are given by
\eql{eq:gDet2}
{
\Phi_k^{\sigma \sigma'} = 
  \begin{bmatrix}
   \phi^{\sigma \sigma'}_{0,0} & \phi^{\sigma \sigma'}_{0,1} & \ldots & \phi^{\sigma \sigma'}_{0,p} & \ldots & \phi^{\sigma \sigma'}_{0,N^{\sigma'}}  \\
   \phi^{\sigma \sigma'}_{1,0} & \phi^{\sigma \sigma'}_{1,1} & \ldots & \phi^{\sigma \sigma'}_{1,p} & \ldots & \phi^{\sigma \sigma'}_{1,N^{\sigma'}}  \\
  \vdots & \vdots &  & \vdots & & \vdots \\
   \phi^{\sigma \sigma'}_{M,0} & \phi^{\sigma \sigma'}_{M 1} & \ldots & \phi^{\sigma \sigma'}_{M,p} & \ldots & \phi^{\sigma \sigma'}_{M,N^{\sigma'}}  \\
   \end{bmatrix}
\,,
}
where $\phi^{\sigma \sigma'}_{i p}$ are  expansion coefficients of orbital $p$, with corresponding spin index $\sigma'$, in terms of the spatial single particle orbital basis functions, $\psi_i$, and the orbitals $\phi^{\sigma \sigma'}$ depend implicitly on the determinant index, $k$.
The propagator
becomes,
 \eql{eq:gBop}
{
B^G(\bm{\sigma}) =
  \begin{bmatrix}
   B^{\uparrow} (\bm{\sigma})&
   B^{soc}(\bm{\sigma})^{\dagger} \\
   B^{soc}(\bm{\sigma}) &
   B^{\downarrow}(\bm{\sigma})
   \end{bmatrix}
\,,
}
where $B^{soc}(\bm{\sigma}) = \mathrm{Exp}[-\tau \hat{W}^{soc}_{\bot}]$ in the absence of spin-dependent electron-electron interactions, and $B^{\uparrow} (\bm{\sigma})$ ($B^{\downarrow} (\bm{\sigma})$)  includes both the usual spin-independent terms, and the axial SOC term.
Thus the random walks in AFQMC  now take place in the space of generalized Slater determinants, with the spin-dependent $B^G(\bm{\sigma})$ operating
directly on GHF determinants. 

Physical observables are computed in terms of the one-body Green's function,
\eql{eq:greensFuncDef}
{
G^{\sigma \sigma'}_{ i j} = \frac{\bra{\Phi_l} \hat{c}^\dagger_{i \sigma} \hat{c}_{j \sigma'}  \ket{\Phi_r} }{\bra{\Phi_l} \Phi_r \rangle}
\,,
}
where the``left'' determinant, $\ket{\Phi_l}$, is a determinant from the trial wavefunction which, in general, is given by $\ket{\Psi_T} = \sum_l C_l \ket{\Phi_l}$ and the ``right'' determinant,  $\ket{\Phi_r}$, is a random walker.
In the case of a single-determinant trial wavefunction, $\ket{\Psi_T} = \ket{\Phi_l}$.
In the absence of SO interactions, the ``spin-flip'' sectors (i.e. $\sigma \neq \sigma'$) are identically zero and only $G^{\uparrow \uparrow}_{ i j}$ and $G^{\downarrow \downarrow}_{ i j}$
are explicitly computed.
When SOC is included we use a combined spatial-spin index $\mu = (i,\sigma)$ such that $c^\dagger_\mu = c^\dagger_{i \sigma}$.
The matrix representation of the Green's function, $G_{ \mu \nu } =  G^{\sigma \sigma'}_{ i j} $, is then given by,
\eql{eq:greensFuncExplicit}
{
   G_{\mu \nu} = \left[  \Phi_r \left(  (\Phi_l)^\dagger \Phi_r  \right)^{-1} (\Phi_l)^\dagger  \right]_{\nu \mu}
\,,
}
where both the left and right determinants are generalized determinants.

The effective system size is doubled 
in a 2-component fully relativistic AFQMC calculation, 
since both spin sectors are explicitly coupled and generalized determinant walkers are used.
Note that, when the SOC terms in the Hamiltonian are zero and the initial population of walkers consists of spin colinear determinants,  both $B^G(\bm{\sigma}) $ and
$\Phi^G_k$  become block diagonal.
Formally, this reduces the method to the spin-free case, as expected. Computationally there would remain an extra cost treating the spin-free case in this 
generic manner, which is removed by recognizing the block diagonal structure as in the standard treatment above in Eqs.~(\ref{eq:sdFac}) and 
(\ref{eq:facBop}). 

\begin{figure}
\begin{center}
\includegraphics[width=0.65\textwidth]{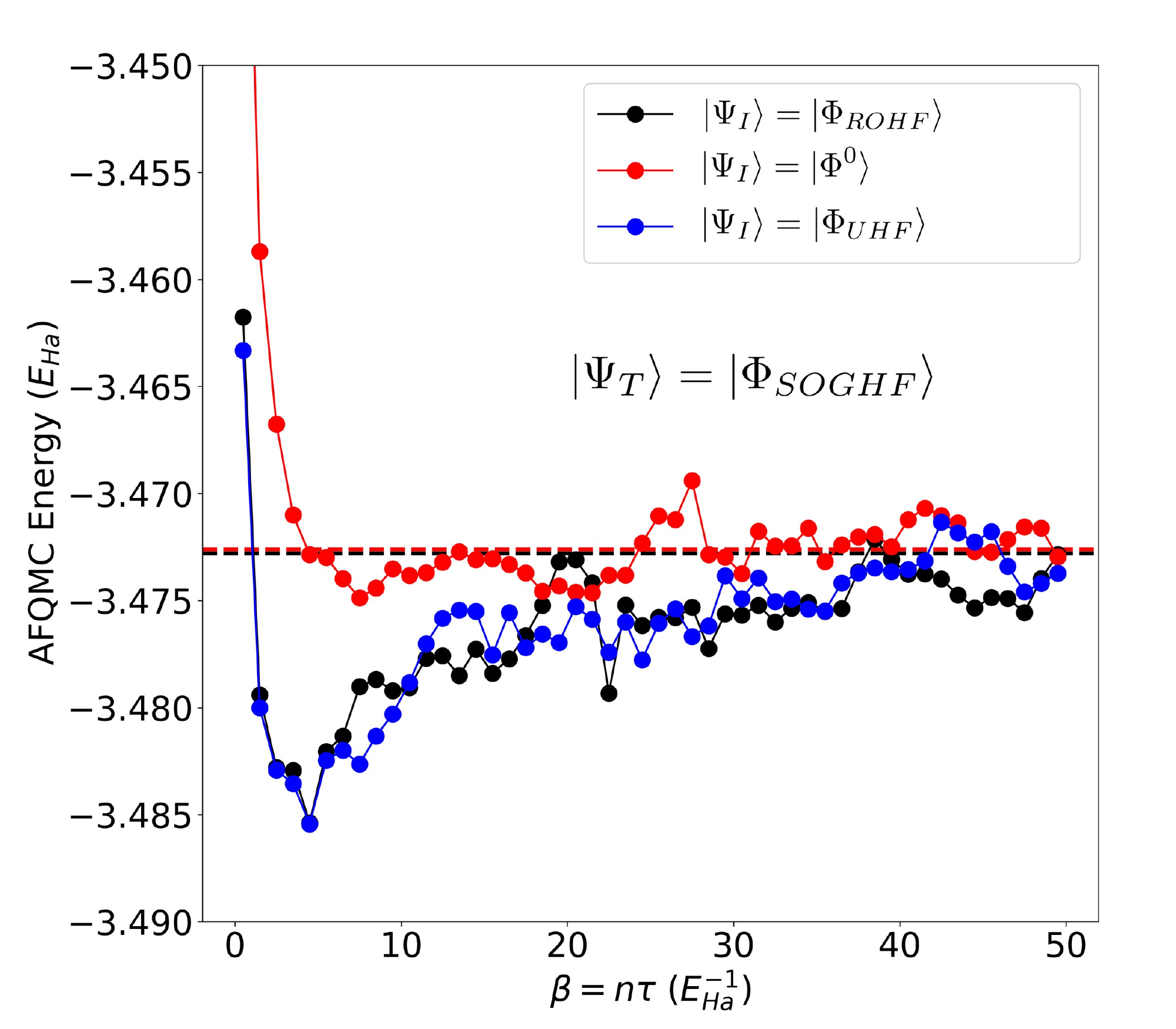}
\caption{\label{fig:spinProj} AFQMC Energy of Pb including SOC: the effect of spin filtration. 
The trial wavefunction is the SO GHF determinant.
The initial walker population is initialized uniformly to a single determinant, $\ket{\Phi_{I}}$, with
each curve corresponding to a particular choice for the initial determinant as described in the text.
Solid lines are a guide to the eye and horizontal dotted lines show the final AFQMC energy using the same total
projection time of $\beta=125.0$ $\textrm{E}^{-1}_\textrm{Ha}$ in all cases.
The dotted lines are nearly indistinguishable since 
the final energies all agree 
to within statistical error bars, which are 
0.6 mE$_{Ha}$ for 
both $\ket{\Phi_{I}} = \ket{\Phi_{UHF}}$ and $\ket{\Phi_{I}} = \ket{\Phi_{ROHF}}$, and 0.3 mE$_{Ha}$ for 
$\ket{\Phi_{I}} = \ket{\Phi^{0}}$, 
the solution to the non-interacting SOC Hamiltonian.
These calculations were performed using the cc-pVQZ-PP basis,
correlating the 6s and 6p electrons.
}
\end{center}
\end{figure}

Much of the technical developments in the spin-free case can be generalized to the present formalism straightforwardly following the above. 
The introduction of the GHF form of the walkers and the spin-dependent propagators does affect the use of
the spin filtration technique \cite{Purwanto2008}, as can be expected.
We illustrate the situation in
Fig.~\ref{fig:spinProj}, which plots the AFQMC energy  for the Pb atom, including SOC, versus total projection time, $\beta$.
We consider three separate AFQMC calculations in which we initialize the projection with different initial populations. 
In all cases the walker has the form of a GHF determinant as is necessary for SOC, and the same trial wavefunction,
the GHF solution in the presence of SOC, is used.
In the first case, the walkers are {\it initialized\/} to a UHF determinant, i.e., in the form of Eq.~(\ref{eq:gDet}) with
$ \Phi_k^{\uparrow \downarrow} =\Phi_k^{\downarrow \uparrow} =0$, and $ \Phi_k^{\sigma\sigma}=\Phi_{UHF}^\sigma$.
A slow convergence is seen in this calculation, which is consistent with the corresponding case in spin-free calculations due to spin contamination.
In the second case, the  walkers are initialized to an ROHF determinant, i.e., $ \Phi_k^{\uparrow \downarrow} =\Phi_k^{\downarrow \uparrow} =0$, and $ \Phi_k^{\sigma\sigma}=\Phi_{ROHF}^\uparrow$ ($=\Phi_{ROHF}^\downarrow$). 
This calculation is a literal application of the spin filtration technique 
for spin-free Hamiltonians, which here exhibits a similar slow convergence to the unfiltered 
calculation with UHF. 
In the third case, we initialize the walkers with the non-interacting solution of the SOC Hamiltonian.
Since $\ket{\Phi^0}$ is an eigenstate of the one-body operator, $\hat{K}$,
 the operation $\mathrm{Exp}[-\tau \hat{K}] \ket{\Phi_0}$ will leave $\ket{\Phi^0}$ unchanged 
 (see Eq.~\ref{eq:HSfull}).
 The remaining factors in the  projection operator (Eq.~\ref{eq:HSfull}) involve only spin-independent operators. 
We see from the results in Fig.~\ref{fig:spinProj} that this calculation 
 shows a better initial relaxation behavior and faster convergence.  
Additionally, it leads to a factor of 2 reduction in the stochastic uncertainty 
in the final AFQMC result compared to the other two calculations.

%
%

\subsection{Computational details}
\label{sec:compDets}

In all calculations, we used small-core Stuttgart PPs~\cite{Peterson2003Br, Peterson2006I, Metz2000,Peterson2003PbBi} obtained from the Stuttgart online database~\cite{StuttgartWeb}.
For ion center $A$, Stuttgart PPs
have the general form,
\eql{eq:relECP}
{
\hat{V}_A^\textrm{PP} = -\frac{Q_A}{r} +  \sum_{l} \sum_{j=| l - 1/2|}^{j=l+1/2} \sum_{k=1}^n B^k_{l j} \mathrm{exp}( -\beta^k_{l j} {r}^2) \hat{\mathcal{P}}_{l j}^A
\,,
}
where $Q_A$ is the ion charge (nuclear charge minus number of core electrons),
$\hat{\mathcal{P}}_{l j}^A$ are projectors for orbital and total angular momentum with respect to the atomic center $A$. 
There are typically $n$=2 or 3 projectors (index $k$ in Eq.~\ref{eq:relECP}) for each angular momentum $l$.
This leads to improved scattering properties and transferability, similar to augmented planewave (PAW), ultrasoft (US) \cite{US_PAW:Kresse1999}, and multiple-projector norm-conserving Vanderbilt-type pseudopotentials \cite{Hamann2013}. 
The parameters $B^k_{l j}$ and $\beta^k_{l j}$  were fit to reference data consisting of all-electron 4-component, multi-configurational Dirac-Hartree-Fock (MCDHF) total energy calculations for about 100 relativistic states with the Breit interaction included perturbatively.
The accuracy of the fit was reported as  $\simeq 10^{-2}$ eV, with maximum deviations of $\simeq 5 \times 10^{-2}$ eV \cite{Metz2000}. 
As improved all-electron relativistic treatments are developed, 
emerging new energy-consistent PPs \cite{Stoll2002} can be easily incorporated into the same framework.

These small core PPs correspond to large active valence spaces. For the Bi atom, for example, the PP treats 23 valence electrons.
For the initial applications of the method in this paper, we used a frozen-core transformation in the target systems to effectively transform these to large-core PPs, similar to previous scalar relativistic only CCSD(T) calculations which also used Stuttgart PPs~\cite{Peterson2003Br,Peterson2006I, Peterson2003PbBi}.
Only the valence s and p electrons are explicitly treated at the many-body level, while
all semi-core electrons are frozen at the 
 scalar relativistic HF/DFT level of theory.
The frozen orbital procedure we used is briefly discussed next.

The frozen orbital transformation we use to obtain an effective Hamiltonian for AFQMC is based on the approximation that the many-body wavefunction can 
be separated into a product form as~\cite{Purwanto2013} 
\eql{eq:frozenOrbSep}
{
\Psi = \mathcal{A} \left(  \Psi_{\mathbb{I}} \otimes \Psi_{\mathbb{A}} \right)
\,,
} where $\mathcal{A}$ is an antisymmetrizer, and $\mathbb{I}$ and $\mathbb{A}$ are  
defined based on mutually exclusive sets of inactive and active orbitals  
chosen from 
physically motivated criteria.
Typically, $\mathbb{I}$ corresponds to the 
 energetic core orbitals, although, in this paper, we treat the energetic semi-core orbitals as ``core''.
The active space Hamiltonian, $ \hat{H}_\mathbb{A}$, is constructed based on the condition that,
\eql{eq:frozenOrbCond}
{
\bra{\Psi} \hat{H} \ket{\Psi} = \bra{\Psi_\mathbb{A}} \hat{H}_\mathbb{A} \ket{\Psi_\mathbb{A}}
\,.
}
Formally, the active Hamiltonian is given by,
\eql{eq:activeH}
{
\hat{H}_\mathbb{A} =  \sum_{\mu \nu \in \mathbb{A} } K_{\mu \nu}  \hat{c}^{\dagger}_\mu  \hat{c}_\nu 
+  \sum_{ \mu \nu \gamma \delta \in  \mathbb{A}} V_{\mu \nu \gamma \delta} \hat{c}^{\dagger}_\mu  \hat{c}^{\dagger}_\nu \hat{c}_\delta \hat{c}_\gamma
+  \sum_{\mu \nu \in \mathbb{A} } V^{\mathbb{I}-\mathbb{A}}_{\mu \nu}  \hat{c}^{\dagger}_\mu  \hat{c}_\nu
+ E_{\mathbb{I}}
\,,
}
where $\hat{K}$ includes all one body terms, including SOC if desired, $\hat{V}^{\mathbb{I}-\mathbb{A}}$ is formally a nonlocal PP which captures the interaction between the active and the inactive electrons,
 $E_{\mathbb{I}}$ is the constant energy contribution due to the frozen electrons including both one- and two-body contributions, and all sums are over orbitals in the active space.
When $\hat{H}$ is based on a small-core PP, Eq.~\ref{eq:activeH} can be thought of as effectively increasing the size of the PP core; however, this approach has improved transferability over a large-core PP with the same active space size, since the semi-core space is built for the specific environment
rather than just the atom.

AFQMC calculations with SOC use GHF 
determinants as a trail wavefunction;
however, the frozen
core procedure above is done with
 scalar relativistic ROHF for Br, I, Pb, and Bi atoms or from scalar relativistic ROKS (LDA) for solid Bi.
The GHF determinants computed for the scalar relativistic Hamiltonian are found to be equivalent to UHF determinants.
All HF/DFT calculations were performed using PySCF unless otherwise stated in the text.
GHF calculations are performed using the same Hamiltonian as AFQMC.

We use the corresponding cc-pVxZ-PP (x=D, T, Q, etc.) basis sets~\cite{Peterson2003PbBi,Peterson2003Br}, which are included in the basis set library of PySCF~\cite{PYSCF2018,PYSCF2020}, and have been optimized for use with the Stuttgart PPs.
In electron affinity calculations, the aug-cc-pVxZ-PP basis set is used.
The cc-pVxZ-PP and aug-cc-pVxZ-PP basis sets are optimized specifically for valence-only correlation and are a natural choice of basis given the effectively large-core PP used.

AFQMC calculations are performed using 20-80 cores clocked at 2.4 GHz with 6.4 GB of memory / core.
Due to the inherently parallel nature of AFQMC, larger calculations than those performed in the present work can easily be performed 
using more compute nodes, and more memory/node.

%
%

\section{Applications}
\label{sec:Apps}

In this section, we illustrate our 
SOC-AFQMC method and implementation with several test applications
in atomic and molecular systems 
as well as a crystalline system.
The post-d elements are known to have strong intrinsic SOC  due to both their high atomic number and 
their open p-shell ground state electron configurations.
For example, the halogens have an ns$^2$np$^5$ ground state configuration corresponding to a $^2\textrm{P}$ L-S term if SOC is neglected.
Including SOC  
splits the configuration into a 4-fold degenerate $^2\textrm{P}_{3/2}$ experimentally observed ground state, consistent with Hund's rules, 
and a 2-fold degenerate $^2\textrm{P}_{1/2}$ excited state where the magnitude of splitting increases with the atomic mass.
As another example, Pb has a large atomic number, 82, and an open shell 6s$^2$6p$^2$ ground state configuration. However, due to very strong SOC, it has a $^3P_2$ experimentally observed ground state~\cite{PbExp_fromMitas}, violating Hund's rules which predict $^3P_0$. 

Similar to Pb, atomic Bi has very strong intrinsic SOC strength due to both its large atomic number, 83, and its ground state electron configuration of $6\textrm{s}^26\textrm{p}^3$.
Therefore, inclusion of relativistic effects, including SOC, is essential to a quantitatively accurate description of Bi.
Crystalline Bi is interesting due to its topological character, which arises from a combination of its lattice structure (described in Section~\ref{sec:Bi}) and strong SOC.
Two-dimensional bilayers of (111) Bi, which resemble the geometry of hexagonal BN, have one-dimensional quantum spin Hall states along the outer edge~\cite{PhysRevLett.97.236805,Drozdov2014} and display higher-order topological states in three-dimensional nanostructures~\cite{Schindler2018}.
This is thus an example in extended systems in which the capability of \textit{ab initio} treatment of SOC and electron interaction on equal footing, with predictive accuracy, is important for future progress.

In this section, we begin with the calculation of the electron affinity of atomic Pb, 
followed by the dissociation energy of the heavy halogen dimers, Br$_2$ and I$_2$, 
and ending with the cohesive energy and equation of state of crystalline Bi.
In all cases, we compare with available experimental data from the literature.

%
%

\subsection{Electron affinity (EA) of lead}
\label{sec:Pb}

\begin{figure}
\begin{center}
\includegraphics[width=0.45\textwidth]{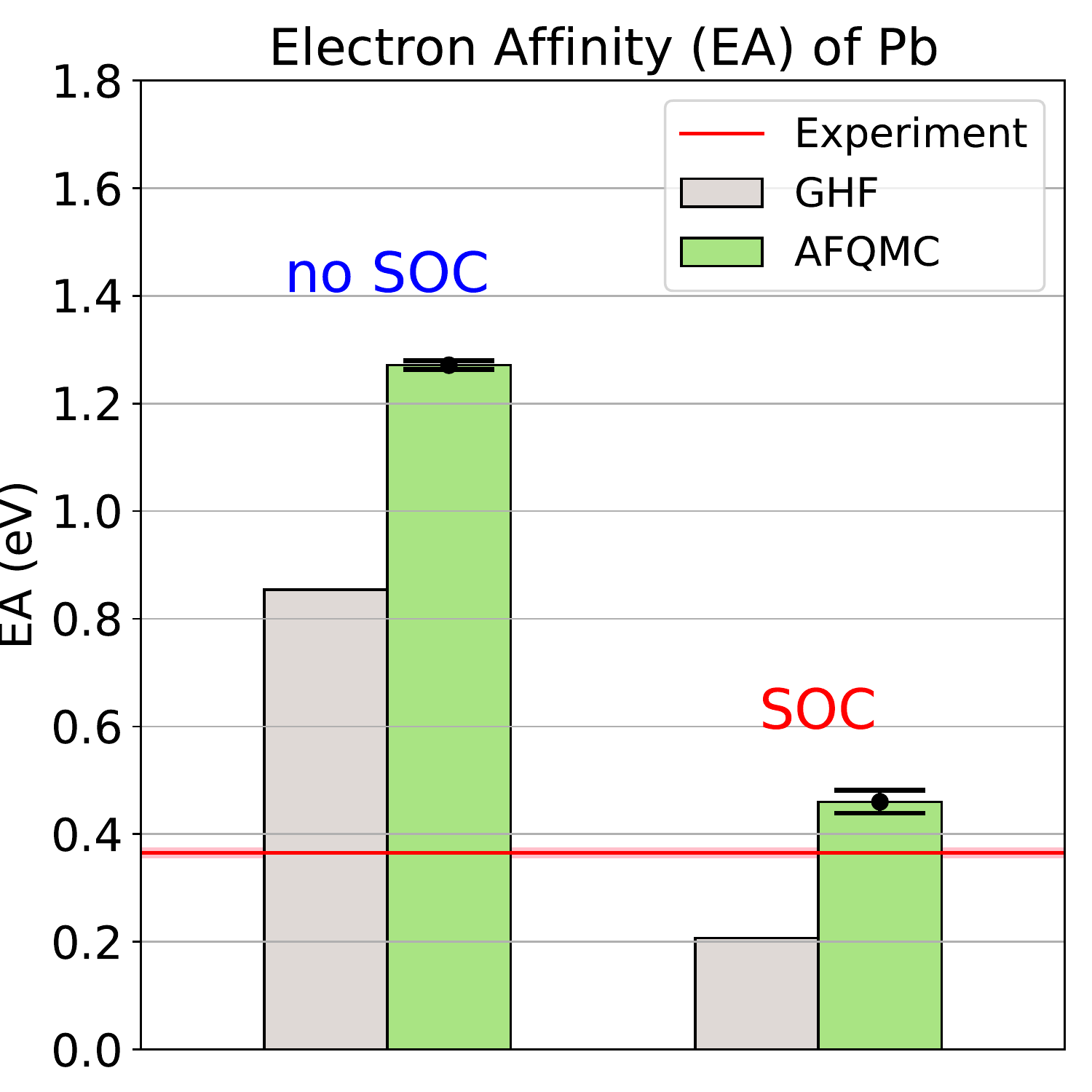}
\caption{\label{fig:Pb_EA} Electron Affinity of Pb.
Gray and green
 bars are, respectively, GHF and AFQMC results.
 The red horizontal line is the experimentally observed electron affinity~\cite{PbExp_fromMitas}, and the width of the line is the experimental uncertainty;
 ``no SOC"  and ``SOC" correspond, respectively, to the scalar relativistic and spin-orbit coupling Hamiltonians.
Calculations were performed using the aug-cc-pVQZ-PP basis.
The AFQMC stochastic uncertainty is displayed as black horizontal bars.
 }
\end{center}
\end{figure}

The EA of Pb is a good test case, because the ground state configurations of the Pb atom and Pb$^-$ ion are qualitatively different and can be
expected to lead to different SOC effects.
Neglecting SOC, the Pb atom has 4 valence electrons in an open-shell 6s$^2$6p$^2$ spin-triplet configuration, 
while the Pb$^-$ ion has 5 valence electrons in the 6s$^2$6p$^3$ spin-quartet configuration with zero orbital angular momentum. 
As a result, SOC is expected to have a large effect on the EA. This is indeed the case, as shown in Fig.~\ref{fig:Pb_EA}.
AFQMC calculations are performed as described in Sec.~\ref{sec:compDets}.
To minimize basis set errors, we used the large aug-cc-pVQZ-PP basis (see Sec.~\ref{sec:2cHam}).
Previous scalar relativistic CCSD(T) calculations indicated an EA difference of only 6 meV between the aug-cc-pVQZ-PP and aug-cc-pV5Z-PP basis sets~\cite{Peterson2003PbBi}.

The large SOC effect arises from the open-shell neutral Pb atom, while the closed-shell 
Pb$^-$ ion, shows relatively small SOC effect.
SOC-AFQMC results, which are shown in Fig.~\ref{fig:Pb_EA} together with that from the corresponding result from the GHF trial wave function used,
is in good agreement with experiment.
Of course, the accuracy of the Stuttgart PP places a limit on the expected agreement between SOC-AFQMC and experiment with typical deviations in the PP spectrum of 0.02 eV, compared to all-electron 4-component reference data, and maximum deviations of up to 0.05 eV which apply separately to the neutral atom and the negative ion.
The discrepancy between SOC-AFQMC and experiment, 0.10(2) eV, is of a similar order of magnitude to this limit, but there 
could be residual discrepancies.
Agreement with experiment can be slightly improved with a larger active space and appropriate core-valence basis.
We tested this by including the 5d$^{10}$ electrons in the active space.
Repeating the frozen-core transformation described in 
Section~\ref{sec:compDets} yields a smaller-core PP, with effective core which is intermediate between the Stuttgart small-core PP, and the effective large-core PP described in the previous section.
For this test, we used the aug-cc-pwCVQZ-PP basis which is optimized for core-valence correlation.
Compared to the aug-cc-pVQZ-PP basis with only 6s and 6p electrons treated using AFQMC,
the EA is decreased by only 0.023(33) eV in the SOC case and by 0.039(21) eV in the scalar relativistic case.

%
%

\subsection{Halogen dimer (Br$_2$, I$_2$)  dissociation energy}
\label{sec:Br2}

The halogens are known to have large SOC strength relative to atoms with a similar atomic number, and are a common test case~\cite{10.1063/1.471636,10.1063/1.5023750,10.1021/acs.jctc.7b00989,2014Br2_4cCI,Sharma2018}.
Here we focus on the dissociation energy of the Br and I dimers, computed as
\mbox{$D_e = 2E_\mathrm{atom} - E_\mathrm{dimer} $},
where $E_{\rm dimer}$ and $E_{\rm atom}$ are the ground-state energies of the dimer and atom, respectively.
The dimer energy
is calculated at the experimentally observed equilibrium bond length, 2.281 \AA for Br$_2$~\cite{BrExp} and 2.666 \AA for I$_2$~\cite{Huber1979}. 
Neglecting SOC, the halogen atoms have a 6-fold degenerate ground state with 
 ns$^2$np$^5$  $^2\textrm{P}$ configuration where n is 4 and 5, respectively, for Br and I. 
Including SOC splits the configuration into a 4-fold degenerate $^2\textrm{P}_{3/2}$ experimentally observed ground state, consistent with Hund's rules, 
and a 2-fold degenerate $^2\textrm{P}_{1/2}$ excited state.
Near equilibrium, the orbital angular momentum of the halogen dimers is mostly quenched, leading to only small SOC effects.
Thus the SOC splitting in the halogen atoms tends to reduce $D_e$.

\begin{figure}
\begin{center}
\includegraphics[width=0.75\textwidth]{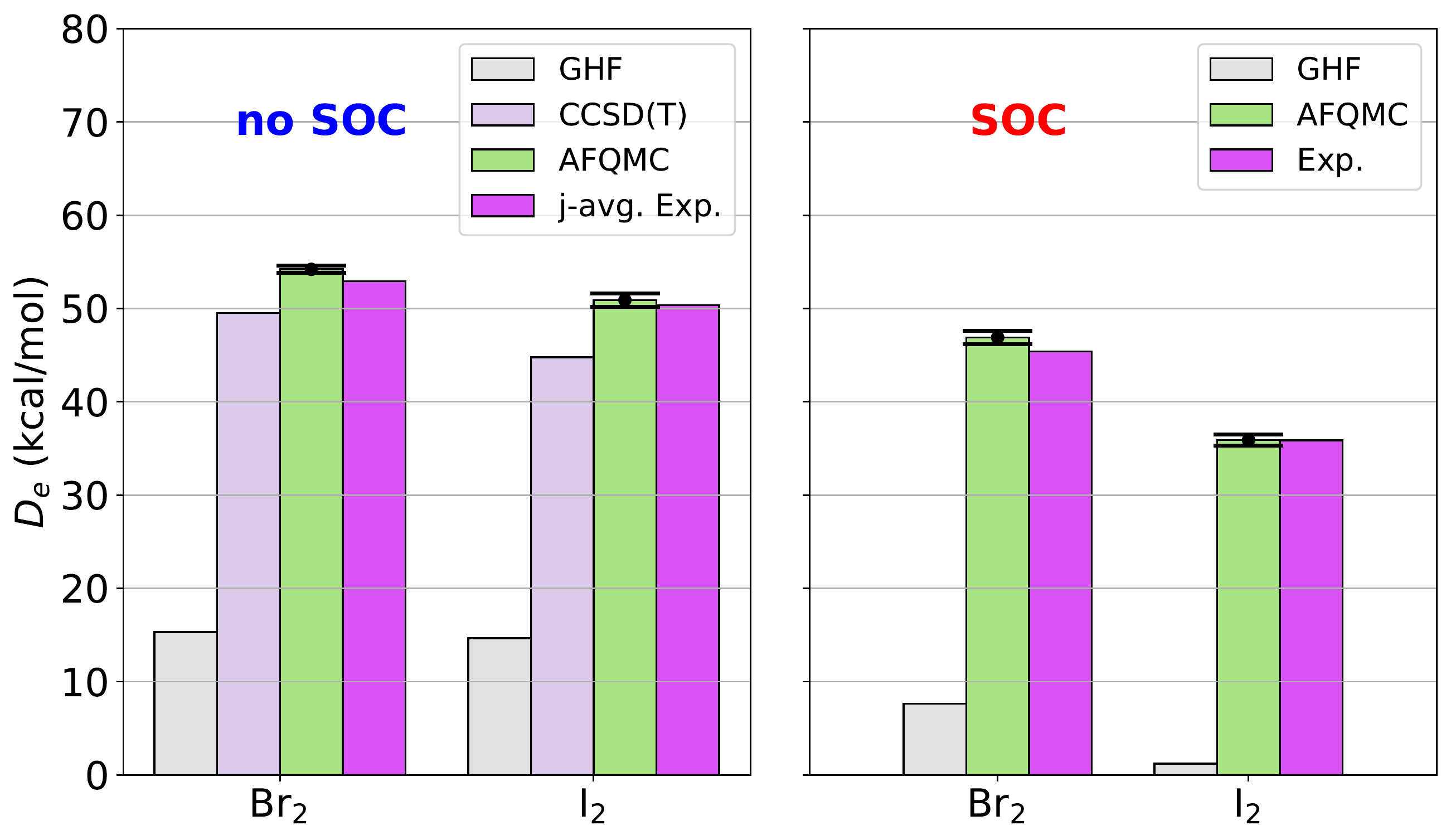}
\caption{\label{fig:Br_De} Dissociation Energy ($D_e$) of Br and I Dimers. 
``no SOC'' and ``SOC'' correspond, respectively, to scalar relativistic calculations without SOC and calculations including SOC.
In the left panel, the experimental reference has been averaged over the ground state j-multiplet~\cite{Peterson2003Br} in order to facilitate comparison with scalar relativistic calculations 
by approximately removing spin-orbit coupling effects.
Errorbars are shown as bold vertical lines with horizontal caps in black and indicate the joint stochastic uncertainty.
All calculations use the cc-pVQZ-PP basis.
}
\end{center}
\end{figure}

Fig.~\ref{fig:Br_De} compares the AFQMC dissociation energy $D_e$ to experiment. 
AFQMC calculations, and the GHF calculations from which the trial wave function is obtained, 
were all performed using
the cc-pVQZ-PP basis.
For comparison, 
scalar-relativistic CCSD(T)~\cite{Peterson2003Br} results using the cc-pVQZ-PP basis are also shown.
(The CCSD(T) result for I$_2$ uses an older form of the Stuttgart PP~\cite{Peterson2003Br} than the one used for AFQMC
here~\cite{Peterson2006I}, but both versions should provide similar $D_e$.)
In the left panel, the bar labeled "j-avg. exp" corresponds to the experimental result in which SOC has been approximately removed by averaging over the observed ground state j-multiplet.
Neglecting SOC, CCSD(T) underestimates the j-averaged experimental result by 3.41 kcal/mol and 5.6 kcal/mol, respecitvely, for Br$_2$ and I$_2$ , and AFQMC overestimates the j-averaged experimental result by 1.2(4) for Br$_2$ and agrees with it for I$_2$ to within the stochastic uncertainty of 0.7 kcal/mol.
With SOC included, AFQMC is within 1.5(7) kcal/mol of the measured value for Br$_2$ and agrees to within the stochastic uncertainty (0.6 kcal/mol) for I$_2$.
GHF greatly underestimates $D_e$ with and without SOC.
According to previous scalar-relativistic CCSD(T) calculations~\cite{Peterson2003Br} which include results up to cc-pV5Z-PP, there are significant residual basis set effects in the cc-pVQZ-PP basis, and
the complete basis set (CBS) limit of $D_e$ for Br$_2$ computed with CCSD(T) is 52.13 kcal/mol.
From this result, we estimate that the AFQMC $D_e$ is increased 
by about 2.6 kcal/mol from the present cc-pVQZ-PP basis to the CBS limit for Br$_2$ and by roughly 2 kcal/mol for I$_2$ by
similar analysis.

Here, we performed somewhat simplified calculations of the $D_e$ of the halogen dimers in order to focus on the methodology.
In a more quantitatively accurate treatment, the small-core Stuttgart PP could be used along with a core-valence basis~\cite{Peterson2010CV},
without the frozen core transformation described in Sec.~\ref{sec:compDets}.
The perfect agreement between AFQMC and experiment might be somewhat  fortuitous here.
For example, we checked that, for the Br atom in the scalar relativistic case, the AFQMC result is improved by the use of a truncated CASSCF trial wavefunction (using 14 orbitals and 7 electrons), with $D_e$ reduced by 2.6 kcal/mol compared with using a GHF trial wavefunction.
For the $D_e$ of Br$_2$ in Fig.~\ref{fig:Br_De},
 the small bias introduced by the simple GHF trial wavefunction and the 
frozen-core and basis set errors likely counteracted each other. 

%
%

\subsection{Crystalline bismuth}
\label{sec:Bi}

As a demonstration of SOC-AFQMC for a crystalline solid, we present calculations of the equation of state (EOS) and cohesive energy of 
Bi
 in the $\alpha$-arsenic A7 structure
 (R$\bar{3}$m, space group no.~166) \cite{Needs1986,Gonze1988,Gonze2007PRB}.
The A7 structure can be visualized as a distorted cubic NaCl structure (both atoms replaced by Bi) stretched in the [111] direction with a
Peierles distortion along the trigonal axis.The crystal is specified by three parameters: the lattice constant $a$, 
rhombohedral angle $\alpha$, and internal coordinate $z$ ($z=0.5$ for the undistorted cubic NaCl structure), which specifies the
distance between the two Bi atoms in the primitive cell \cite{Gonze2007PRB}.
The AFQMC total energy, with and without SOC, was calculated  
as a function of the lattice constant,
holding $\alpha = 57.37^\circ$ and $z=0.468$ fixed at the experimentally observed values~\cite{Gonze2007PRB}.

In order to focus on the methodology, we use a simplified treatment of Bi.
AFQMC calculations were performed in an 8-atom supercell (distorted NaCl conventional cubic cell) with
periodic boundary conditions, using 
the cc-pVDZ-PP basis set. 
As described in the previous section, the Stuttgart small-core PP was used to 
eliminate core electrons, and the frozen orbital method was used to additionally freeze higher-lying semi-core electrons, 
based on spin-free LDA Kohn-Sham (KS) orbitals, 
resulting in an active space containing five valence electrons per atom.
The trial wave function was obtained from a GHF calculation.
The AFQMC results here are not expected to vary with respect to the choice of the trial wave function.
In the scalar relativistic case, 
we found that HF and LDA trial wave functions produced 
AFQMC energies which agreed to within 2 meV,
similar to previous results in other sp-bonded systems.

We applied several, essentially independent, corrections 
to better approach the CBS and the thermodynamic limit in the solid.
First, 1- and 2-body finite size corrections~\cite{KweeFS} were computed using Quantum Espresso~\cite{QEsp1,QEsp2,QEsp3} with a converged 12x12x12 Monkhorst-Pack (MP) grid 
~\cite{Gonze2007PRB}.
Large core Hartwigsen-Goedecker-Hutter (HGH) pseudopotentials~\cite{HGH1998}, which have 5 valence electrons  per atom for Bi, similar to our AFQMC treatment, were used to compute finite size corrections.
Finite size corrections were computed using the scalar relativistic Hamiltonian, 
and the correction was applied to both the scalar and SOC AFQMC result.
Second, a simple basis set incompleteness correction was included 
by comparing AFQMC calculations in the primitive cell using  cc-pVDZ-PP and cc-pVTZ-PP basis sets.
These corrections were computed separately for the scalar relativistic and SOC cases.

\begin{figure}
\begin{center}
\includegraphics[width=0.9\textwidth]{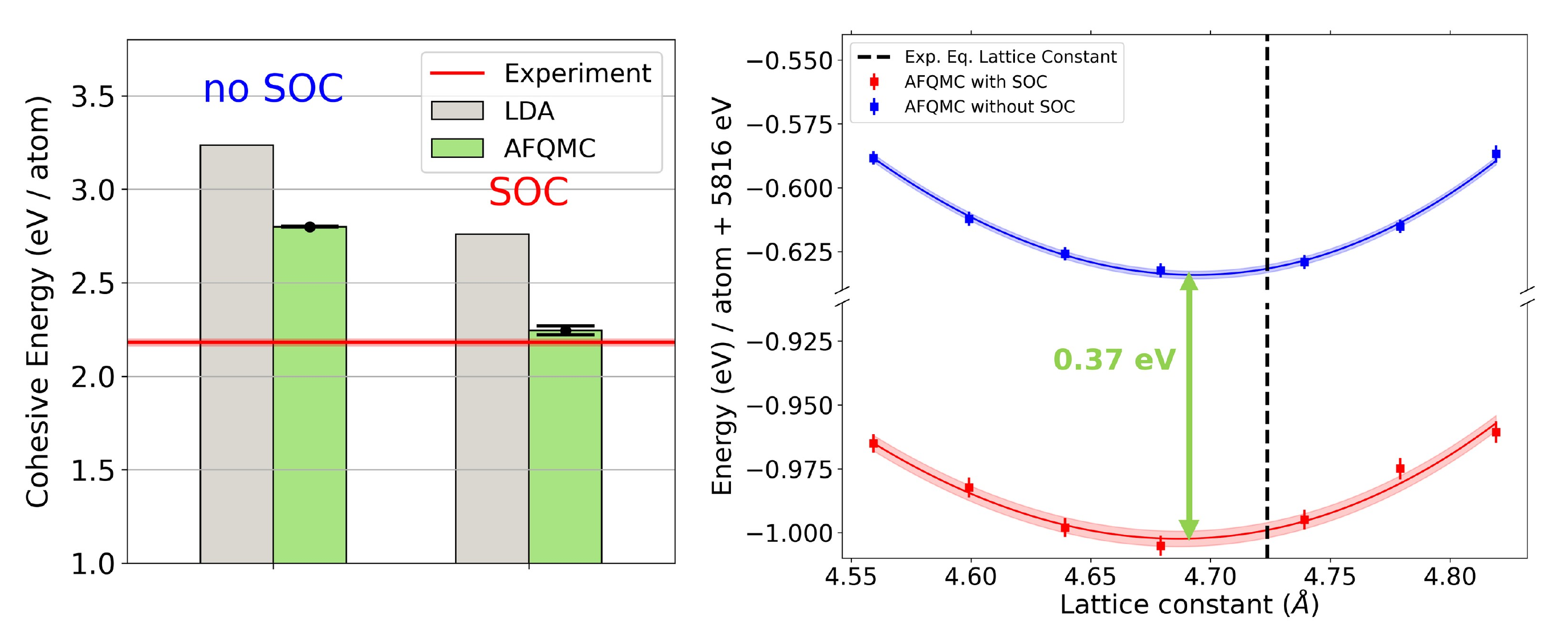}
\caption{\label{fig:Bi_Ec} (left) Cohesive energy of solid Bi in the A7 structure.
The experimentally measured cohesive energy~\cite{LBNL1977} is plotted as a shaded red horizontal line whose width
represents the experimental uncertainty.
AFQMC stochastic uncertainty is indicated by error bars. 
LDA results are from D\'{\i}az-S\'anchez \textit{et al.} (2007b).
(right) AFQMC equation of state (EOS) of A7 Bi with and without SOC.
Only the lattice constant was varied, while the rhombohedral angle $\alpha$
and internal coordinate $z$ were held fixed at the experimentally observed value.
Shaded lines indicate the combined stochastic and fitting errors. The experimental equilibrium lattice constant $a_0$ is indicated by
the vertical dashed line.
}
\end{center}
\end{figure}

The left panel of Figure~\ref{fig:Bi_Ec}
compares the size- and basis set-corrected AFQMC cohesive energies \mbox{$E_{\rm coh} = E_\mathrm{atom}-E_\mathrm{solid}$}
with LDA~\cite{Gonze2007PRL} and experimental~\cite{LBNL1977} values.
The crystalline ground state energy per 
atom, $E_\mathrm{solid}$, was obtained from a quadratic fit of the 
EOS (right panel of Fig.~\ref{fig:Bi_Ec}).
The inclusion of SOC is clearly essential for quantitative agreement with experiment.
With SOC, $E^\mathrm{AFQMC}_{\rm coh} = 2.245(24)$~eV, which overestimates experiment by only 0.065(24) eV,  
while LDA overestimates it by 0.58 eV.
Without SOC, the discrepancy increases to 0.62 and 1.08 eV, respectively for AFQMC and LDA . 
Note that 
the {\it difference} between E$_{\rm coh}$ computed from LDA and AFQMC is
dependent on SOC treatment (0.437(3) vs.~0.515(24) eV), indicating 
that correlation and spin-orbit effects are
entangled, which is somewhat surprising in an s-p bonded material, but is consistent with the very strong intrinsic spin-orbit coupling of Bi.
While the agreement between the present AFQMC calculations and experiment is encouraging, a more thorough comparison 
would require better estimates of $E_{\rm coh}$'s approach to the basis set and thermodynamic limits,
using a larger basis set, with larger supercells and more twist-angles in AFQMC.
 
The EOS of Bi, computed both with and without SOC, is plotted in the right panel of Figure~\ref{fig:Bi_Ec}.
The inclusion of SOC is seen to shift the A7 Bi EOS curve downward by -0.37 eV, which would increase $E_{\rm coh}$ by itself, 
compared to the spin-free case. 
However, the ground state of the isolated Bi atom with SOC is lowered even more, by $\sim -0.92 eV$.
The combined SOC effects result in a net lowering of $E_{\rm coh}$ by $\sim -0.55 eV$, 
bringing the AFQMC prediction into good agreement with experiment,  as depicted in the left panel of Fig.~\ref{fig:Bi_Ec}.
This demonstrates that an accurate treatment of SOC is necessary in both the atomic and solid state 
systems to compute 
 $E_{\rm coh}$ with quantitative accuracy.
The predicted equilibrium lattice constant $a_0$, to within error bars, is the same for the scalar relativistic and SOC EOS, 
deviating from experiment by $\sim 0.6$\,\%.
Previous LDA results\cite{Gonze2007PRB}, where the A7 structure was allowed to fully relax, found that the predicted $a_0$ in the scalar case was reduced 
by 0.042 \AA~ compared to the SOC result, which is essentially identical to our AFQMC value.
In our calculations, as mentioned, the rhomohedral angle $\alpha$ and the internal coordinate $z$ 
were held fixed at the experimentally observed values.
Allowing the lattice to relax at the AFQMC level of theory may lead to slightly different results.

%
%

\section{Conclusion}
\label{sec:Discussion}

We have incorporated explicit, non-perturbative treatment of spin-orbit coupling into \textit{ab initio} AFQMC
calculations, and demonstrated its use to treat test systems including atoms, molecules, and solids.
AFQMC including SOC produced accurate results consistent with its performance in spin-free \textit{ab initio} Hamiltonians. 
For the EA of Pb, AFQMC agrees with experiment to within 0.1 eV.
AFQMC including SOC predicts the correct dissociation energy to within 1.5(7) kcal/mol and 
0.6 kcal/mol for Br$_2$ and I$_2$ respectively.
In solid Bi, AFQMC yields a 
cohesive energy which is within 
0.065(33) eV of the experimental value,  where the error bar accounts for both the stochastic uncertainty of AFQMC and the experimental uncertainty.

In this work, 
the treatment of the test systems is somewhat simplified, in order to focus on the methodology.
For Br, I, and Bi atoms and molecules, larger basis sets can be used for a more precise extrapolation to the CBS limit.
In solid Bi, a better treatment would involve twist-averaged boundary conditions and/or larger supercells.
The quality of trial wave function, which we took from GHF calculations, controls the bias introduced by the phaseless approximation in AFQMC.
Single-determinant trial wave functions are often sufficient for many AFQMC calculations; however, sometimes multideterminant wave functions are needed for reaching chemical accuracy (for example in transition metal containing systems)~\cite{Purwanto2009_C2,Purwanto2014Cr,Shee2019}.
Efficient implementations of multideterminant trial wavefunctions with a sublinear scaling in the number of determinants have been developed~\cite{QMCPack2020,Shi2021,Mahajan2021}.
It would be interesting to test the performance of multideterminant trial wave functions which are computed with the explicit inclusion of SOC.

While relativistic PPs including SOC were used in the present work, any two-component form of the relativistic Hamiltonian, with or without SOC, can be used with no modification to the AFQMC implementation.
In principle, even spin-dependent electron-electron interaction terms could be included at the AFQMC level of theory; in most cases, the effect is small and 
it is unlikely that including such terms would produce significantly better results than the use of relativistic PPs which account for the Breit interaction in the fitting reference data.
The SOC operator is highly local and, therefore, can be incorporated straightforwardly into local embedding AFQMC~\cite{Eskridge2019}, which 
provides a natural tool 
to study spatially large systems containing only a few heavy ions that can 
display important SOC effects.
For example, in molecular magnets, there may be only a handful of lanthanide or actinide ions embedded in a large organic molecule.
It is also straightforward to use relativistic PPs to selectively include SOC only in the desired ions while treating all other atoms with scalar relativistic, or even non-relativistic, Hamiltonian terms.

\section*{Acknowledgements}
\label{sec:ack}

This work is supported by DOE grant number DE‐SC0001303.
The Flatiron Institute is a division of the Simons Foundation.
The authors acknowledge William \& Mary Research Computing for providing computational resources and/or technical support that have contributed to the results reported within this paper.
URL: https://www.wm.edu/it/rc

\bibliographystyle{phaip.bst}
\bibliography{afqmc-soc.bib}

\end{document}